\documentclass[superscriptaddress,nobibnotes,amsmath,amssymb,notitlepage,twocolumn,prl,longbibliography]{revtex4-1}

\usepackage{bm,mathptmx,braket}

\usepackage{graphicx,color,hyperref}
\usepackage{bm}
\usepackage[caption=false]{subfig}
\hypersetup{colorlinks=true, linkcolor=blue, citecolor=blue, urlcolor=blue} 

\graphicspath{{./img/}}

\newcommand{\beq}[1]{\begin{equation}\label{#1}}
\newcommand{\eep}{\;.\end{equation}}
\newcommand{\eec}{\;,\end{equation}}
\newcommand{\eeq}{\end{equation}}

\newcommand*\dd{\mathop{}\!\mathrm{d}} 

\newcommand{\om}{\omega}

\newcommand{\sect}[1]{\vspace{0.3em}{\it #1.}---}

\DeclareMathAlphabet{\mathcal}{OMS}{cmsy}{m}{n} 

\renewcommand{\vec}[1]{{\bf #1}}

\newcommand{\kv}{\vec{k}}

\begin{document}

\title{Probing tensor monopoles and gerbe invariants in three-dimensional topological matter}

\newcommand{\TCM}{{Theory of Condensed Matter Group, Cavendish Laboratory, University of Cambridge, J.\,J.\,Thomson Avenue, Cambridge CB3 0HE, United Kingdom}}

\newcommand{\UoM}{Department of Physics and Astronomy, University of Manchester, Oxford Road, Manchester M13 9PL, United Kingdom}

\newcommand{\Coimbra}{{CFisUC, Department of Physics, University of Coimbra, Rua Larga, 3004-516 Coimbra, Portugal}}

\newcommand{\Dublin}{{School of Theoretical Physics, Dublin Institute for Advanced Studies,
10 Burlington Road, Dublin D04 C932, Ireland}}


\author{Wojciech J. Jankowski}
\email{wjj25@cam.ac.uk}
\affiliation{\TCM}

\author{Robert-Jan Slager}
\affiliation{\UoM}
\affiliation{\TCM}

\author{Giandomenico Palumbo}
\email{giandomenico.palumbo@gmail.com}
\affiliation{\Coimbra}
\affiliation{\Dublin}

\date{\today}

\begin{abstract}
We show that momentum-space tensor monopoles corresponding to nontrivial vector bundle generalizations, known as bundle gerbes, can be realized in bands of three-dimensional topological matter with nontrivial Hopf invariants. We provide a universal construction of tensor Berry connections in these topological phases, demonstrating how obstructions therein lead to $\mathbb{Z}$-quantized bulk magnetoelectric and nonlinear optical phenomena in the presence of spacetime inversion ($\mathcal{PT}$) symmetry. We then pinpoint that these quantum effects are supported by intraband and interband torsion leading to nontrivial Dixmier--Douady classes in most known Hopf phases and in more general topological insulators realizing gerbe invariants falling beyond the tenfold classification of topological phases of matter. We furthermore provide an interacting generalization upon introducing many-body gerbe invariants by employing twisted boundary conditions. This opens an avenue to study gerbe invariants realized through higher-dimensional charge fractionalizations that can be electromagnetically probed.
\end{abstract} 

\maketitle

\sect{Introduction} A key manifestation of topology in physics is embodied by quantization conditions \cite{Nash1988}. Fundamentally, physical topological quantizations can be understood in terms of monopole structures. Examples range from the quantization of charge in particle physics in the form of Abelian Dirac monopoles \cite{Preskill1984} to the quantization of the Hall conductivity in condensed matter physics~\cite{Thouless1982, Haldane1988}. Monopoles are generally inherently related to topological invariants in quantum field theory, in which there exist different kinds such as the Yang monopole in five dimensions \cite{Yang1978} and the 't Hooft-Polyakov monopole associated with non-Abelian gauge theories \cite{Hooft1974,Polyakov1974}.
Besides Abelian and non-Abelian vector gauge fields, tensorial gauge fields play a central role in high-energy physics, such as in gravity and string theory with relevant applications in condensed matter physics. In particular, the tensorial character of gauge fields allows for a coupling to extended and inhomogeneous force fields, such as strings~\cite{Kalb1974, tong2012}, which provides for an emergence of exotic physical phenomena inaccessible to point-like objects. For instance, suitable symmetric tensor gauge fields have been shown to describe fractons, which are dual to elasticity supporting topological defects in the form of disclinations or dislocations~\cite{Pretko2018,Beekman20171}, and higher-spin field theory \cite{Fronsdal1978}, a lower-dimensional version of which has recently been employed to study the p-atic phases of the fractional quantum Hall effect \cite{Bergshoeff2024}.

In string theory, antisymmetric tensor gauge fields known as Kalb--Ramond fields $B_{\mu \nu}$~\cite{Kalb1974} allow for an existence of monopoles sourcing a higher-order field strength $\mathcal{H}_{\mu \nu \rho}$, known as tensor monopoles \cite{Nepomechie1985,Orland1982}, which have been experimentally realized in several synthetic matter systems~\cite{Tan2021, Cappellaro2022, Mo2025}. At the same time, the higher flux $\mathcal{H}_{\mu \nu \rho}$ is also manifested via a kinematic term in the Kalb--Ramond Lagrangian, ${\mathcal{L} = -\frac{1}{4} \mathcal{H}_{\mu \nu \rho} \mathcal{H}^{\mu \nu \rho}}$, analogously to the Maxwell Lagrangian encompassing electrodynamics \cite{Henneaux1986} and in the equations of~motion associated to the topological \textit{BF} theories \cite{Cattaneo1995,Hansson2004,Cho2011, Palumbo2025-2}. Furthermore, in addition to its natural relation to Abelian gauge theories, tensorial flux $\mathcal{H}_{\mu \nu \rho}$ has also been shown to be directly related to the fully antisymmetric components of the torsion tensor in torsionful gravity~\cite{Scherk1974,tong2012}.

Importantly, the mathematical roots of antisymmetric tensorial gauge theories are encoded in gerbe bundles \cite{Murray1996,Carey2000}, generalizations of vector bundles that naturally deal with cohomology of higher-forms and differential topology in higher dimensions. The relevance of gerbes acquired a recent interest in topological band theory~\cite{Palumbo2018, Palumbo2019, Zhu2020,Zhu2021, Palumbo2021}, and a particularly intensified focus within tensor network constructions for characterizing interacting topological phases of matter~\cite{Ohyama2024, Shiozaki2023, Ohyama2024_2, Ohyama2025, Sommer2025, Qi2025}. One of the main results of this approach is an identification of novel topological classes and invariants known as Dixmier--Douady classes and numbers, respectively \cite{Johnson2003,Szabo2012}. These Dixmier--Douady numbers are defined as integers in odd dimensions, differently from Chern numbers defined in even dimensions.

The main goal of this Letter is twofold: (i) we provide a construction of tensor Berry connections~\cite{Palumbo2018, Palumbo2019, Zhu2020,Zhu2021, Palumbo2021}, i.e., momentum-space Kalb--Ramond fields,  for fermionic condensed matter systems that could support not only stable topological invariants, but also delicate {topology}~\cite{Nelson2022}.  Most interestingly, we show that this also leads to interacting generalizations. In addition, (ii) we identify their observable manifestations provided by tensor monopole quantizations embodied by nontrivial Dixmier--Douady classes in topological gapped phases, such as Hopf insulators \cite{Moore2008,
Kennedy2016, Schuster2019,  Unal2019, Lim2023, Jankowski2024PRB} with underlying quaternionic structures~\cite{Unal2020}, which we show to translate into their linear and nonlinear electromagnetic responses in the presence of spacetime inversion ($\mathcal{PT}$) symmetry. Importantly, we retrieve these quantizations in delicate topological insulators, i.e., falling beyond the tenfold classification based on $K$-theory~\cite{Kitaev2009, FreedMoore, Slager2017}, and unstable against the hybridizations with additional trivial bands~\cite{Nelson2022}, showing that the tensorial gauge structures are intrinsic to homotopy-classified phases of matter, i.e., phases with topological invariants defined only under finite multiband partitionings~\cite{Bouhon2020}. As such, we propose distinct ways to probe nontrivial bundle gerbe structures and tensorial monopoles in topological phases with nontrivial Dixmier--Douady classes, such as Hopf insulators that realize delicate topology~\cite{Nelson2022}. In this Letter, we thus demonstrate physically observable manifestations of tensor connections in a class of magnetic topological insulators with broken time-reversal symmetry, and with particle conservation symmetry.

\begin{figure}[t!]
\centering
\includegraphics[width=\columnwidth]{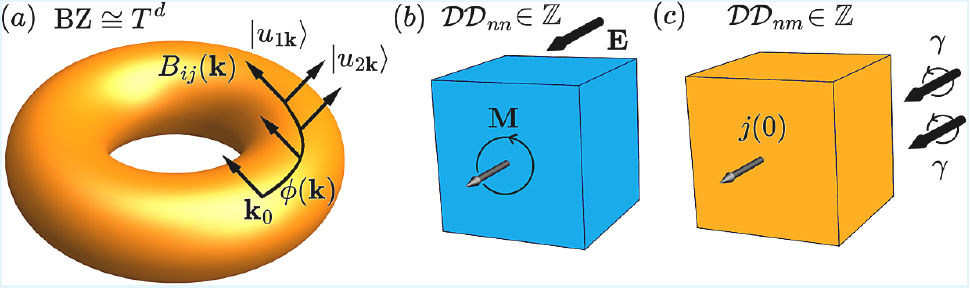}
\caption{Responses from nontrivial gerbe bundles in topological \mbox{insulators}.
{ (a)} Construction of the pseudoscalar $\phi(\kv)$ and tensor connection $B_{ij}(\kv)$ from Bloch states $\ket{u_{1\kv}}, \ket{u_{2\kv}}$ and $\phi(\kv)$ over $d$-dimensional Brillouin zone (BZ) realizing topology of a torus~($T^d$). { (b)} Magnetoelectric response linear in electric field $\vec{E}$, probing intraband ${\mathcal{DD}_{nn} \in \mathbb{Z}}$ invariants.
{ (c)} Nonlinear shift current $j(0)$ response to a pair of circularly polarized photons $\gamma$, quadratic in electric field, probing interband $\mathcal{DD}_{nm} \in \mathbb{Z}$ ($n \neq m$) gerbe invariants.
}
\label{Fig1}
\end{figure}

\sect{Bundle gerbe invariants} We now detail the construction of tensorial Berry connections amounting to gerbe bundle invariants, which we further retrieve as observable in certain topological insulators beyond the tenfold classification~\cite{Kitaev2009}. We begin with standard vector Berry connections central to Hall phenomena. In condensed matter physics, the Hall phenomena in two dimensions are induced by the Berry curvature flux in the space of momenta~\cite{Thouless1982, Niu1985},
\beq{}
    F^n_{xy} = \partial_x A^n_y - \partial_y A^n_x,
\eeq
where $A^n_i = \text{i} \braket{u_{n\kv}| \partial_i u_{n\kv}}$ is the vectorial Berry connection, $\kv$~is a single-particle momentum, $\ket{u_{n\kv}}$ is the $n$-th Bloch band eigenvector, and we use ${\partial_i \equiv \partial_{k_i}}$ for brevity. Formally, $A^n_i$ is a connection one-form over a line bundle of eigenvectors $\ket{u_{n\kv}}$ over Brillouin zone (BZ) in $k$~space, and can be seen as the momentum-space analog of the electromagnetic potential.
In a similar way, an existence of a momentum-space version of the Kalb--Ramond field, dubbed tensor Berry connection $B_{ij}$, with ${i,j=x,y,z}$, has been proposed~\cite{Palumbo2018, Palumbo2019, Zhu2020, Zhu2021, Palumbo2021}, which generates the following fluxes of the momentum-space tensorial gauge fields in three dimensions
\beq{eq:KRflux}
    \mathcal{H}_{xyz} = \partial_x B_{yz} + \partial_y B_{zx} + \partial_z B_{xy}.
\eeq
More formally, $B_{ij}$ can be thought of as a connection of the bundle gerbe [see Sec.~I of the Supplemental Material (SM) for details~\cite{SI}], which we here retrieve on gauging fluxes $F_{ij}$ with pseudoscalar fields $\phi$~\cite{Palumbo2019}. As a central result, we will construct the tensor connection $B_{ij}$ from the Wilson lines of the Berry connections of Bloch states. 

Crucially, the monopoles in tensor Berry connection fields defined over bundle gerbes provide for nontrivial Dixmier--Douady ($\mathcal{DD}$) invariants, as an obstruction to Gauss--Ostrogradsky divergence theorem,
\beq{}
    \mathcal{DD}_{nm} = -\frac{1}{4\pi^2} \int_{\text{BZ}} \dd^3 \kv~\mathcal{H}^{nm}_{xyz},
\eeq
with ${n,m}$ the band indices, which is a higher-dimensional analog of the Chern class quantization as an obstruction to Stokes's theorem,
\beq{}
    C_{n} = \frac{1}{2\pi} \int_{\text{BZ}} \dd^2 \kv~F^{n}_{xy}.
\eeq
It is well-known that the Chern class underlies the quantization of the Hall conductivity ${\sigma_{xy}=\frac{e^2}{h} \sum_n C_n}$ supported by individual occupied bands $n$~\cite{Thouless1982}. On the contrary, here, we construct $\mathcal{H}^{nm}_{xyz}$ from $B^{nm}_{ij}$, following Eq.~\eqref{eq:KRflux}, with ${B^{nm}_{ij} \equiv \phi_{nm} \mathcal{F}^{mn}_{ij}}$, where ${\mathcal{F}^{nm}_{ij} = \partial_i A^{nm}_j - \partial_j A^{nm}_i}$, with ${A^{nm}_i = \braket{u_{n\kv} |\partial_{k_i}u_{m\kv}}}$ the derivative part of the non-Abelian Berry curvature flux. As such, by construction, ${\partial_k\mathcal{F}^{nm}_{ij} +\partial_i\mathcal{F}^{nm}_{jk}+ \partial_j\mathcal{F}^{nm}_{ki}=0}$, which shows that tensor monopoles require singular features in $\phi_{nm}(\kv)$. However, differently from the previous gerbe constructions~\cite{Palumbo2018, Palumbo2019, Zhu2020, Zhu2021, Weisbrich2021, Palumbo2021}, and centrally to this work, here we define the pseudoscalar fields as follows,
\beq{eq:phi}
    \phi_{nm}(\kv) \equiv \int^{\kv}_{\kv_0} d{k'}^i~ A^{nm}_i(\kv'),
\eeq
which formally is equivalent to a Wilson line from point ${\vec{k}_0=\vec{0}}$ over the momentum space [see Fig.~\ref{Fig1}(a)]. In the above, we assume Einstein summation convention, as in the rest of this work. {By default, we choose the shortest path, i.e., geodesic, over the parameter space as an integration contour, see Fig.~\ref{Fig1}(a)}. We demand that the pseudoscalar fields are real by construction, which for ${n \neq m}$ must be enforced with an additional symmetry, e.g., spacetime inversion symmetry ($\mathcal{PT}$), such that the Hamiltonian and the eigenvectors can be written in real gauge~\cite{Zhao2017, Bouhon2020}. We note that the above construction resembles the dimensional reduction of one-form gauge fields that allows obtaining axion fields in quantum field theory \cite{Reece2025,Palumbo2025}. In fact, such pseudoscalar fields can be seen as momentum-space versions of real-space axions. The pseudoscalar phases $\phi_{nm}(\textbf{k})$ play a key role in realizing the nonvanishing ${\mathcal{DD} \in \mathbb{Z}}$ invariants. In particular, if ${\phi_{nm}(\textbf{k}) = \text{const}}$, then ${\mathcal{H}_{xyz}=0}$, identically. The physical interpretation of $\phi_{nm}(\textbf{k})$ as accumulated geometric phases amounts~to positional displacements of the wave packets formed in bands $n$ and $m$, upon changing their center-of-mass momentum from $\textbf{k}_0$ to momentum $\textbf{k}$, consistently with the formalism of Ref.~\cite{Gao2019}.

Notably, shifting the origin ${\vec{k}_0=\vec{0}}$ to another base point $\vec{k}'_0$ {collinear with $\vec{k}$ and $\vec{k}_0$}, transforms the pseudoscalars as ${\phi_{nm}(\kv) \rightarrow \phi_{nm}(\kv) + \phi^0_{nm}}$, with ${\phi^0_{nm} \equiv \int^{\kv'_0}_{\kv_0}  d{k'}^i A^{nm}_i(\kv')}$. Here, the path between $\textbf{k}_0$ and $\textbf{k}'_0$ is the shortest path across the BZ. These can be viewed as gauge transformations of the gerbe two-form connections,
\beq{eq:trans}
    B^{nm}_{ij} \rightarrow B^{nm}_{ij} + \partial_i \xi^{nm}_j - \partial_j \xi^{nm}_i \equiv B^{nm}_{ij} + \Lambda^{nm}_{ij} ,
\eeq
with gauge vectors ${\xi^{nm}_i = \phi^0_{nm} A^{mn}_i}$, defining an exact two-form $\Lambda^{nm}_{ij}$, which is divergenceless and cannot contribute to the flux $\mathcal{H}^{nm}_{ijk}$, nor to the $\mathcal{DD}$ invariants. Moreover, specifically for ${\phi_{nn} \equiv \int^{\kv}_{\kv_0} d{k'}^i A^{n}_i(\kv')}$, we admit a local transformation ${A^{n}_i(\kv) \rightarrow A^{n}_i(\kv) + \partial_i \alpha^n(\kv)}$ on transforming the Bloch states ${\ket{u_{n\kv}} \rightarrow e^{\text{i}\alpha_n(\kv)}\ket{u_{n\kv}}}$, ${\phi_{nn}(\kv) \rightarrow \phi_{nn} (\kv) + \alpha_n (\kv)}$. Under the non-Abelian transformations with matrices $U$, ${\ket{u_{m\kv}} \rightarrow [U]_{mn} \ket{u_{n\kv}}}$, the pseudoscalar matrix, ${[\Phi]_{nm} \equiv \phi_{nm}}$, transforms as ${\Phi \rightarrow U \Phi U^\dagger}$, consistently with the vector of non-Abelian Berry connection matrices, ${[\vec{A}]_{nm} \equiv (A^{nm}_x, A^{nm}_y, A^{nm}_z)}$, ${\vec{A} \rightarrow U \vec{A} U^\dagger}$ \cite{Palumbo2021}. $U$ is a general unitary transformation that can depend on $\textbf{k}$. By construction, we can choose a gauge: ${A^{nm}_i(\kv) = \partial_i\phi_{nm}(\kv)}$. Upon such particular gauge fixing for non-Abelian Berry connection, ${\mathcal{H}^{nn}_{xyz} = \varepsilon^{abc} A^n_a \partial_b A^n_c}$, or ${\mathcal{H}^{nm}_{xyz} = \varepsilon^{abc} A^{nm}_a \partial_b A^{mn}_c}$, which are equal to Abelian Chern-Simons forms amounting to quantized Hopf and real Hopf indices~\cite{Moore2008, Unal2020, Lim2023, Jankowski2024PRB} in topological insulators, when integrated over single-particle eigenstates across BZ. Importantly, although the numerical value of $\mathcal{H}_{xyz}$ can coincide with the Abelian Chern-Simons upon gauge fixing, $\mathcal{H}_{xyz}$ is an independent gauge invariant observable quantity, which can be viewed as a contribution of individual quasiparticle with momentum $\textbf{k}$ to the magnetoelectric gyration, analogously to the Hall drift contributions from the Berry curvature $F^n_{ij}$ in the Hall current. {Notably, the gauge-dependent Chern--Simons form gives rise to a gauge-invariant tensorial flux only when taken in a~special, real gauge. Appealing to the real gauges allows isolating interband coherences underpinned by the non-Abelian Berry connections, which physically underlie the introduced \mbox{constructions} of tensor Berry connections and fluxes.}

In the following, we show that the $\mathcal{DD}$ invariants, including intraband gerbe invariants $\mathcal{DD}_{nn}$, quantize geometric contributions to magnetoelectric responses, providing for $\mathbb{Z}$ quantizations in realistic three-dimensional topological crystallites [see Fig.~\ref{Fig1}(b)]. Furthermore, we show that the more general interband $\mathcal{DD}$ gerbe invariants $\mathcal{DD}_{nm}$ quantize nonlinear second-order integrated shift responses to circularly polarized light [see Fig.~\ref{Fig1}(c)].

\begin{figure}[t!]
\centering
\includegraphics[width=\columnwidth]{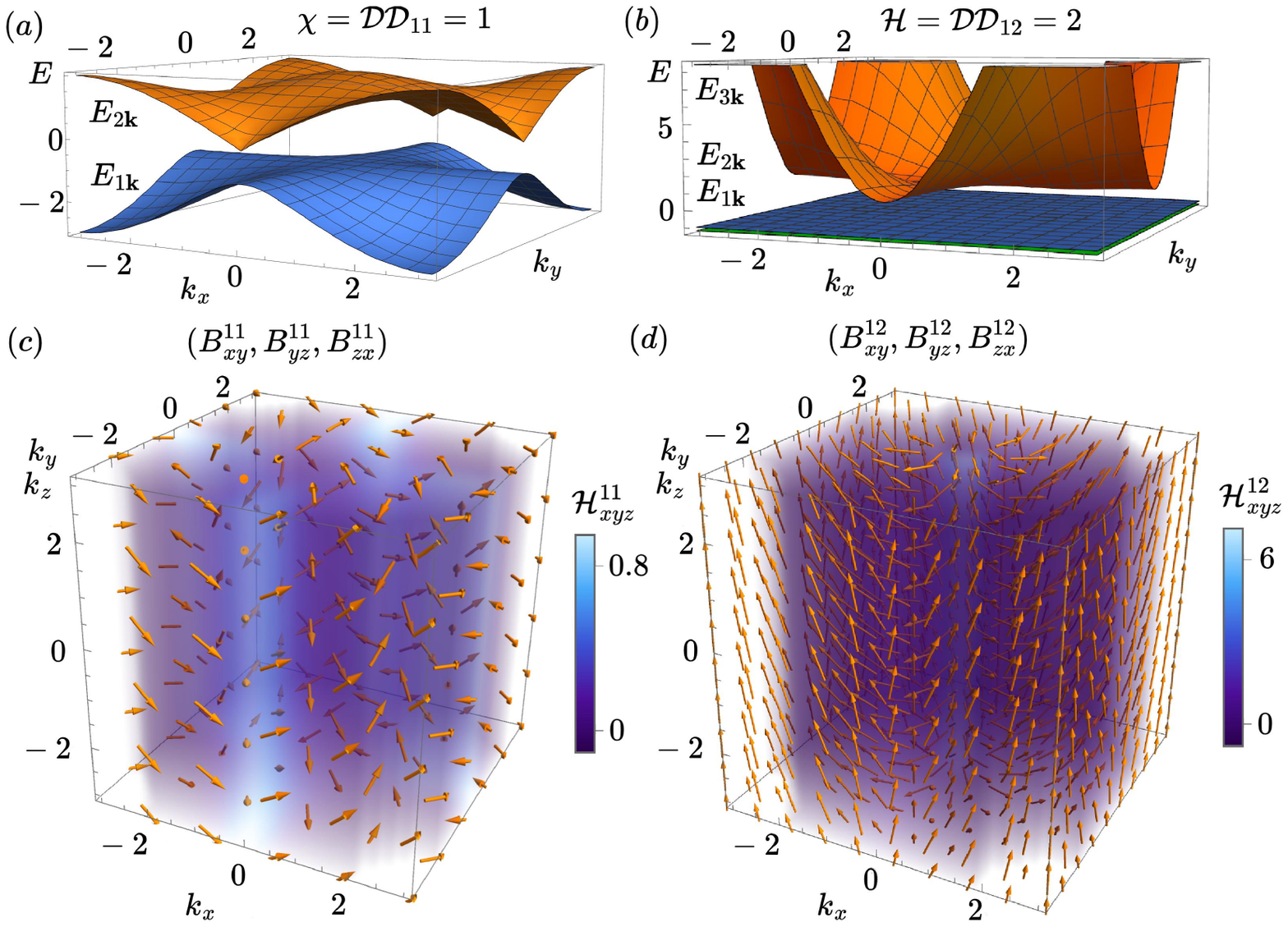}
\caption{\textbf{Tensor monopoles in topological insulators with Hopf indices.}
{\bf (a)} Band structure of two-band Hopf insulator realizing intraband ${\mathcal{DD}_{11} \in \mathbb{Z}}$ invariant at ${M=1}$ and {\bf (b)} three-band real Hopf insulator realizing interband ${\mathcal{DD}_{12} \in \mathbb{Z}}$ invariant at ${M=3/2}$. The bands are shown at momentum ${k_z=0}$. 
{\bf (c)} Tensor monopole of momentum-space Kalb--Ramond field $B^{11}_{ij}$ yielding flux $\mathcal{H}^{11}_{ijk}$ in Hopf insulator with ${\chi = 1}$.
{\bf (d)} Tensor monopole of Kalb--Ramond field $B^{12}_{ij}$ introducing an interband torsional flux $\mathcal{H}^{12}_{ijk}$ in real Hopf insulator with topological index ${\mathcal{H} = 2}$.
}
\label{Fig2}
\end{figure}

\sect{Gerbes and magnetoelectric effects} Here, we retrieve the gerbe quantization by the $\mathcal{DD}$ invariants in magnetoelectric responses~\cite{Qi2008, Malashevich2010, Malashevich2010_1, Qi2011RMP, Shiozaki2013}. The magnetoelectric tensor $\alpha_{ij}$ yielding orbital magnetization currents ($M_i$) from electric fields~$E_j$, ${M_i = \alpha_{ij} E_j}$, consists of topological and nontopological terms (see Sec.~II of SM~\cite{SI}), where the topological term dominates in tenfold-classified topological materials~\cite{Malashevich2010_1}. We~find that the trace of topological magnetoelectric coupling amounts to the higher Berry curvature and $\mathcal{DD}$ invariants,
\beq{eq:ME}
\begin{split}
    \text{Tr}~\alpha^\text{top}_{ij} = -\frac{e^2}{8 \pi^2 h c} \int_{\text{BZ}} \dd^3 \kv~\sum_{n,n'} \mathcal{H}^{nn'}_{xyz} = \frac{e^2}{2 h c} \sum_{n,n'} \mathcal{DD}_{nn'}.
\end{split}
\eeq
Here, the Kalb--Ramond flux over occupied bands, ${\mathcal{H}^{nn'}_{xyz} = \varepsilon^{ijk} \mathcal{T}^{nn'}_{ijk}}$, admits intraband torsion $\mathcal{T}^{nn}_{ijk}$, when ${n=n'}$, which we introduce as a generalization of the interband torsion $\mathcal{T}^{nn'}_{ijk}$~\cite{Ahn2022, Jankowski2024PRB} (see Sec.~III of SM~\cite{SI}). In particular, in a~Hopf insulator~\cite{Moore2008},
\beq{}
    \text{Tr}~\alpha^\text{top}_{ij} = \frac{e^2}{2 h c} \mathcal{DD}_{11} = \frac{e^2}{2 h c} \chi,
\eeq
with ${\chi \in \mathbb{Z}}$ the Hopf invariant, consistently with the $\mathbb{Z}$-quantized Chern-Simons form, ${\theta_\text{CS} = \pi \chi}$~\cite{Alexandradinata2021}. We identify that irrespectively of the value of $\theta_\text{CS}$, defined up to $2\pi$ as an angle~\cite{Alexandradinata2021}, the gauge-invariant gerbe flux $\mathcal{H}^{nn'}_{xyz}$ yields a $\mathbb{Z}$-magnetoelectric effect in the Hopf insulator. Analogously, the $\mathbb{Z}$-magnetoelectric effect arises in the chiral three-dimensional insulators with winding numbers in the chiral symmetry class AIII~\cite{Shiozaki2013, Wang2015}. In fact, in these strong topological \mbox{insulators}, the $\mathbb{Z}$-quantization of the magnetoelectric effect can be also formally explained by employing the gerbe formalism, as~the~three-dimensional winding number coincides with the $\mathcal{DD}$~invariant~\cite{Palumbo2019, Shiozaki2026}.

\begin{figure}[t!]
\centering
\includegraphics[width=\columnwidth]{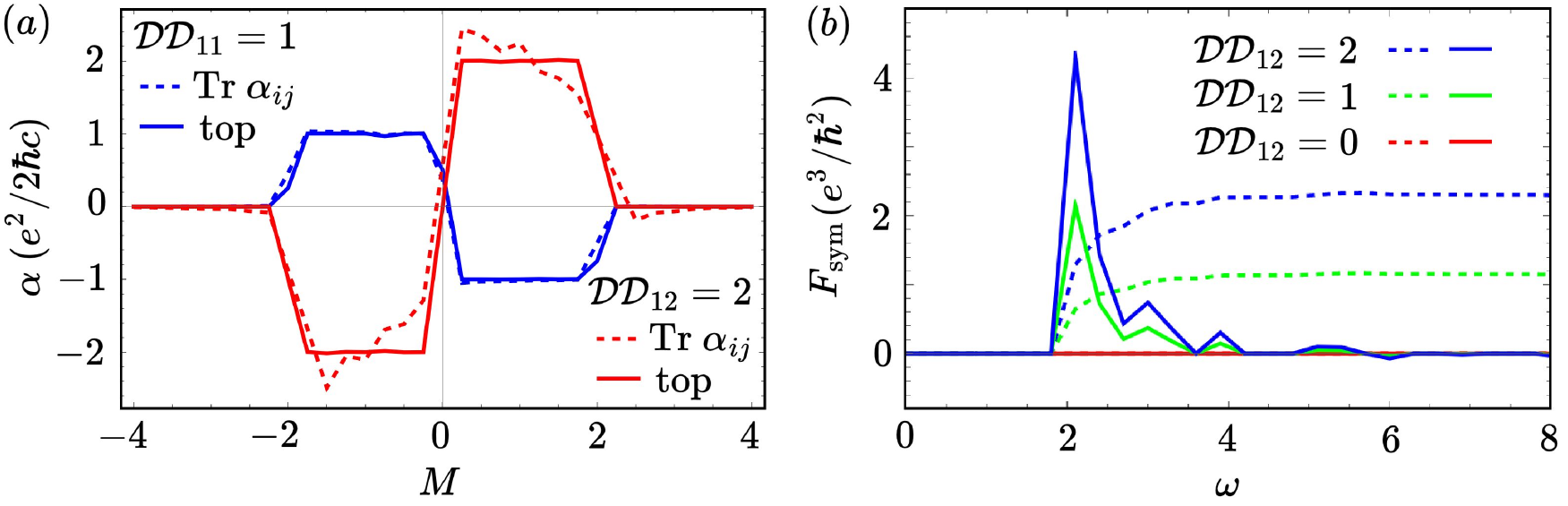}
\caption{\textbf{Quantized responses and gerbe invariants.}
{\bf (a)} Topological ($\text{top}$) magnetoelectric susceptibility $\text{Tr}~\alpha^\text{top}_{ij}$ (\text{solid}) and total ($\text{Tr}~\alpha_{ij}$) magnetoelectric susceptibility (\text{dashed}) in Hopf and real Hopf insulators, as a function of topological mass parameter $M$. Topological phases with ${\mathcal{DD}_{11}=1}$, ${\mathcal{DD}_{12}=2}$ correspond to ${|M|<2}$. The nontopological magnetoelectric response is dominated by the topological contribution.
{\bf (b)} Integrated shift response against cutoff frequency ${\om=\om_\text{max}}$ (\text{dashed}) for different interband gerbe invariants $\mathcal{DD}_{nm}$ in three-band real Hopf insulators with lower two bands occupied. The frequency resolution (\text{solid}) of the integrated response indicates dominant contributions from the Kalb--Ramond flux $\mathcal{H}^{nm}_{xyz}$ due to fermions photoexcited at frequencies directly above the band gap.
}
\label{Fig3}
\end{figure}

In two-band Hopf insulator realizations~\cite{Moore2008, Kennedy2016}, with an occupied topological band ${n=1}$, the invariant reads ${\mathcal{\chi} = -\frac{1}{4\pi^2} \int_\text{BZ} \dd^3 \kv~ [\vec{A}_{11} \cdot (\nabla_\kv \times \vec{A}_{11})] \in \mathbb{Z}}$, which represents the nontrivial homotopy elements ${\pi_3 [S^2] \cong \mathbb{Z}}$, mathematically corresponding to distinct Hopf fibrations: ${S^3 \rightarrow S^2}$~\cite{Moore2008}. We show the corresponding Hopf bands in Fig.~\ref{Fig2}(a), tensor monopole [Fig.~\ref{Fig2}(c)], and topological magnetoelectric response quantized by the gerbe invariant against the corrections from nontopological responses, as a function of topological mass parameter $M$ in Fig.~\ref{Fig3}(a). For further details on the models, topological classifications, and underlying gerbe structures, see the Supplemental Material (SM)~\cite{SI}.

Moreover, we find that the topological magnetoelectric response is quantized by an~\textit{interband} $\mathcal{DD}$ invariant in three-band real Hopf insulators with an occupied pair of topological bands $\{ \ket{u_{1\kv}}$, $\ket{u_{2\kv}} \}$,
\beq{}
    \text{Tr}~\alpha^\text{top}_{ij} = \frac{e^2}{2 h c} \mathcal{DD}_{12} = \frac{e^2}{2 h c} \mathcal{H},
\eeq
where we define ${\mathcal{H} = -\frac{1}{4\pi^2} \int_\text{BZ} \dd^3 \kv~ [\vec{A}_{12} \cdot \vec{Eu}_{12}] \in \mathbb{Z}}$ which is quantized under $\mathcal{PT}$ spacetime inversion symmetry~\cite{Jankowski2024PRB}, with ${\vec{Eu}_{12} \equiv \nabla_\kv \times \vec{A}_{12}}$ the Euler curvature~\cite{Ahn2019, Bouhon2020ZrTe, bouhon2023quantumgeometry}. Under the quantizing symmetry, ${\mathcal{H} \in \mathbb{Z}}$ furthermore represents distinct nontrivial homotopy elements ${\pi_3 [S^2] \cong \mathbb{Z}}$~\cite{Jankowski2024PRB}. We stress that this quantized magnetoelectric response, which we here show to arise from the gerbe invariant, was not identified in Ref.~\cite{Jankowski2024PRB}, which introduced the three-band Hopf--Euler band topologies, where a trivial $\theta$~angle, ${\theta_\text{CS} = 0~(\text{mod}~2\pi)}$, was only recognized. We show the corresponding real Hopf bands in Fig.~\ref{Fig2}(b), the tensor monopole [Fig.~\ref{Fig2}(d)], and the magnetoelectric response against the topological mass $M$ for phases with a distinct topological index $\mathcal{DD}_{12}$ [Fig.~\ref{Fig3}(a)]; see SM (Sec.~I)~\cite{SI} for details on the models. The identification of the Hopf insulator as a~condensed matter realization of a~momentum space tensor monopole, 
\beq{}
\chi = -\frac{1}{4\pi^2} \int_\text{BZ} \dd^3 \kv~\mathcal{H}_{xyz} \in \mathbb{Z},
\eeq
is a central result of this Letter. The integer character of ${\chi = \mathcal{DD}_{11}}$ can be magnetoelectrically probed, e.g., within the previously proposed general experimental setup~\cite{Shiozaki2013}. Furthermore, the $N$-band generalization of a Hopf insulator~\cite{Lapierre2021} also realizes a tensor monopole, upon a reformulation of the invariant as a~tensor flux, which we explicitly demonstrate in the SM~\cite{SI}.

\sect{Gerbes in quantized shift responses} In the following, we show how the gerbe structure associated with interband $\mathcal{DD}_{nm}$ invariants (${n \neq m}$) can be probed with nonlinear second order optical responses. Inspired by high energy physics results in real spacetime \cite{Scherk1974}, we here employ the momentum-space correspondence between the gerbe flux and the totally antisymmetric part of the torsion, namely ${\mathcal{H}^{mn}_{ijk} =  \mathcal{T}^{mn}_{[ijk]}}$, with $\mathcal{T}^{mn}_{ijk}$ interband torsion, which captures virtual transitions between bands ${n,m}$ through additional bands $p$~\cite{Ahn2020, Ahn2021, Jankowski2024PRL}. We first recast the torsional sum rule of Ref.~\cite{Jankowski2024PRL} for circular shift photoconductivities $\sigma_{ijk}^{\text{shift,C}}(\om)$~\cite{Ahn2020} that determine geometric dc photovoltaic current density responses ${j_i(0)=\sigma_{ijk}^{\text{shift,C}}(\om)E_j(\om)E_k(-\om)}$ from electronic positional shifts~\cite{1982JETP, Sipe1993, Morimoto2016, Cook2017, alexandradinata2022topological, Jankowski2024PRL} induced by optical electric fields $E_j(\om)$, which we further detail and comment in Sec.~III of SM~\cite{SI},
\\
\beq{eq:shift}
\begin{split}
    F_\text{sym} &\equiv \int^{\om_{\text{max}}}_0 \dd \om~ [\sigma_{xyz}^{\text{shift,C}}(\om) + \sigma_{yzx}^{\text{shift,C}}(\om) + \sigma_{zxy}^{\text{shift,C}}(\om)] \\&= -\frac{e^3}{4\pi^2 \hbar^2} \int_{\text{BZ}}  \dd^3 \kv \sum_{m,n} f_{nm} \mathcal{H}^{mn}_{xyz}.
\end{split}
\eeq
$\om_{\text{max}}$ is an energy cutoff targeting optical transitions between all single-particle states in occupied/unoccupied bands $n/m$. The filling factor differences ${f_{nm} \equiv f_n - f_m}$ are fixed on setting a zero-temperature limit, with ${f_n = 1}$ for occupied and ${f_m = 0}$ for unoccupied bands, which results in a general torsional quantization from fully occupied bands
\beq{eq:shiftDD}
    F_\text{sym} = \frac{e^3}{\hbar^2} \sum_{n \neq m} \mathcal {DD}_{nm}.
\eeq
We show the integrated shift response against the cutoff frequency $\om_\text{max}$ for phases with distinct topological index $\mathcal{DD}_{12}$ in Fig.~\ref{Fig3}(b); see SM (Sec.~I)~\cite{SI} for details on the models. While a similar physical quantization condition was retrieved in $\mathcal{PT}$-symmetric three-band and four-band models~\cite{Jankowski2024PRL, Jankowski2024PRB}, the more general gerbe structures providing the quantization as tensor monopoles related to momentum-space tensor Berry connections, as well as their additive character underpinned by Eq.~\eqref{eq:shiftDD}, have not been uncovered before. This moreover provides a generalization to arbitrary $N$-band topologies realizing $\mathcal{DD}$ invariants, beyond the previous four-band and three-band realizations of topological insulators with Hopf indices~\cite{Lim2023, Davoyan2024, Jankowski2024PRL, Jankowski2024PRB}, as the sum over ${n,m}$ with $n$ occupied and $m$ unoccupied bands is otherwise unconstrained.

\sect{Many-body generalization}
Importantly, the discussed linear and nonlinear electromagnetic responses can also be utilized to study interacting systems. For example, optical probes, such as circular dichroism, have been employed to study interacting two-dimensional topological systems, such as fractional Chern insulators~\cite{Repellin2012}.
Generalizing our results to interactions in three-dimensional topological systems central to this work, necessarily involves many-body
$\mathcal{DD}$ invariants.
Hence, as an outlook, we briefly detail how the $\mathcal{DD}$ invariants can be generalized to interacting systems with twisted boundary conditions analogous to the construction of many-body Chern invariants~\cite{Thouless1982}. We impose twisted boundary on a many-body wavefunction $\ket{\psi(\bm{\theta})}$ with twist angles ${\bm{\theta} = (\theta_x, \theta_y, \theta_z)}$  acting as twisted boundary conditions across linear system size lengths $L_i$~\cite{Niu1985}, ${\psi(\{x_i + L_i\}) = e^{\text{i} \theta_i} \psi(\{x_i\})}$, where $\{x_i\}$ are coordinates of the single particles constituting the many-body wavefunction in many-body particle coordinate representation, ${\psi(\{x_i\}) = \braket{\{x_i\}|\psi(\bm{\theta})}}$. We define the many-body connection, ${A_i(\bm{\theta}) \equiv \text{i}\braket{\psi(\bm{\theta})|\partial_{\theta_i} \psi(\bm{\theta})}}$, and many-body flux, ${F_{ij}(\bm{\theta}) = \partial_{\theta_i} A_j(\bm{\theta}) - \partial_{\theta_j} A_i(\bm{\theta})}$, alongside a \mbox{pseudoscalar},
\beq{eq:int}
    \phi(\bm{\theta}) \equiv \int^{\bm{\theta}}_{\bm{\theta}_0} \dd {\theta}'_i~A_i(\bm{\theta}').
\eeq
with a gauge freedom of choosing $\bm{\theta}_0$. The \mbox{combination} of flux and pseudoscalar defines a tensor connection ${B_{ij}(\bm{\theta}) = \phi(\bm{\theta})F_{ij}(\bm{\theta})}$. Analogously to the single-particle~case, the many-body tensor connection $B_{ij}(\bm{\theta})$ transforms as ${B_{ij}(\bm{\theta}) \rightarrow B_{ij}(\bm{\theta}) + \Lambda_{ij}(\bm{\theta})}$, with divergenceless two-form $\Lambda_{ij}(\bm{\theta})$, under the shift of the initial integration limit specifying the zero of twisted boundary conditions ${(\bm{\theta} = 0)}$ within the many-body pseudoscalar definition. The many-body tensor connection determines the tensorial flux,
\beq{}
    \mathcal{H}_{xyz}(\bm{\theta}) = \partial_{\theta_x} B_{yz}(\bm{\theta}) + \partial_{\theta_y} B_{zx}(\bm{\theta}) + \partial_{\theta_z} B_{xy}(\bm{\theta}),
\eeq
and integrates to a many-body $\mathcal{DD}_{\text{MB}}$ invariant,
\beq{}
    \mathcal{DD}_\text{MB} \equiv -\frac{1}{4\pi^2} \iiint \dd \theta_x \dd \theta_y \dd \theta_z \mathcal{H}_{xyz}(\bm{\theta}) \in \mathbb{Z}.
\eeq
On one hand, in the presence of short-range interactions, and no ground-state degeneracy, we expect that this invariant remains quantized, similarly to the quantization of the first Chern number in two-dimensional interacting systems~\cite{Hastings,Bachmann}. On the other hand, in the case of a $q$-degenerate ${(q \in \mathbb{N})}$ ground state, we expect the $\mathcal{DD}_\text{MB}$ invariant to be fractionalizable (${\mathcal{DD}_\text{MB} = 1/q}$), analogously to the many-body Chern invariant~\cite{Niu1985,Hastings}. The fractionalization could possibly be probed with the discussed linear and nonlinear electromagnetic effects that target the gerbe invariants. The realizations and detailed calculations of fractional $\mathcal{DD}_\text{MB}$ invariant in specific three-dimensional interacting and topologically ordered phases will be addressed in future work.

\sect{Discussion and conclusion} We demonstrate that gerbe bundle invariants provided by tensor gauge fields associated with tensor monopoles can be probed in three-dimensional fermionic phases of matter such as Hopf insulators and other topological phases with nontrivial $\mathcal{DD}$ invariants. We retrieve probes of these structures in integer magnetoelectric responses targeting the intraband gerbe invariants, as well as nonlinear optical responses probing interband torsion and gauge-invariant Kalb--Ramond fluxes $\mathcal{H}^{nm}_{ijk}$ selectively at the momentum space sections where the transitions at the photon frequency $\om$ occur, see Fig.~\ref{Fig3}. In doing so, we provide a unique construction of tensor Berry connections in topological insulators, beyond the conventional Berry connections, showing that momentum-space Kalb--Ramond fields can also be manifestly realized by free fermionic topological phases.

Crucially, in this Letter, we have shown \mbox{$\mathbb{Z}$-magnetoelectric} phenomena in condensed matter systems supported by tensor Berry connections and $\mathcal{DD}$ \mbox{invariants}, apart from significantly generalizing the quantized shift \mbox{effects} beyond the realms of previous works.

Our findings are applicable to three-dimensional magnetic materials, as well as to quantum simulation setups, which proved successful in simulating matter with nontrivial Hopf indices~\cite{Schuster2021}. As in the case of interactions fractionalizing Chern invariants, we derive the many-body generalization and predict the fractionalization of the $\mathcal{DD}$ invariant, here defined under twisted boundary conditions, to arise from many-body ground-state degeneracies. The provided electromagnetic probes could allow to pinpoint a possible existence of fractional topological insulators characterized by the many-body $\mathcal{DD}$ invariant via magnetoelectric and optical phenomena originally retrieved here for integer quantizations under the noninteracting setups.

\sect{Acknowledgments} We thank Aris Alexandradinata, Zory Davoyan, Duncan Haldane, Hans Hansson, Charles Kane, Mandela Mehraeen, and Eugene Mele for helpful discussions. W.J.J.~acknowledges funding from the Rod Smallwood Studentship at Trinity College, Cambridge. R.-J.S. acknowledges funding from a~New Investigator Award, EPSRC Grant No. EP/W00187X/1, a EPSRC ERC underwrite Grant No.  EP/X025829/1, and a~Royal Society exchange Grant No. IES/R1/221060, as well as Trinity College, Cambridge. This work was financed through national funds by FCT -- Fundação para a Ciência e Tecnologia, I.P. in the framework of the project UID/04564/2025, with DOI identifier 10.54499/UID/04564/2025.

\bibliography{references}

\end{document}


\title{{\textsc{supplemental material}} \\  Probing tensor monopoles and gerbe invariants in three-dimensional topological matter}

\newcommand{\TCM}{{Theory of Condensed Matter Group, Cavendish Laboratory, University of Cambridge, J.\,J.\,Thomson Avenue, Cambridge CB3 0HE, UK}}

\newcommand{\UoM}{Department of Physics and Astronomy, University of Manchester, Oxford Road, Manchester M13 9PL, UK}

\newcommand{\Coimbra}{{CFisUC, Department of Physics, University of Coimbra, Rua Larga, 3004-516 Coimbra, Portugal}}

\newcommand{\Dublin}{{School of Theoretical Physics, Dublin Institute for Advanced Studies,
10 Burlington Road, Dublin D04 C932, Ireland}}


\author{Wojciech J. Jankowski}
\email{wjj25@cam.ac.uk}
\affiliation{\TCM}

\author{Robert-Jan Slager}
\affiliation{\UoM}
\affiliation{\TCM}

\author{Giandomenico Palumbo}
\email{giandomenico.palumbo@gmail.com}
\affiliation{\Coimbra}
\affiliation{\Dublin}

\date{\today}

\date{\today}

\maketitle

\begin{widetext}

\tableofcontents 

\newpage

\section{More details on gerbes}\label{app::A} The gerbe connection can be constructed from the contact structure constituted by a triplet of scalar fields $(\phi^\chi_1, \phi^\chi_2, \phi^\chi_3)$ over a base space, which in the context of this work is the $k$-space~\cite{Palumbo2019}. We set the flavours $\chi$ to run over the orbital basis $\chi = A, B, \ldots$. The $\phi^\chi_a$ fields, which encode the contact structure, define a tensor connection~\cite{Palumbo2019},
\beq{}
    B^{nm}_{ij} =  \frac{\text{i}}{3} \varepsilon^{abc} \sum_\chi \phi^\chi_a \partial_i \phi^\chi_b  \partial_j \phi^\chi_c,
\eeq
with $a,b,c = 1,2,3$. We choose $\phi^\chi_1 = u^\chi_n$, $\phi^\chi_2 = (u^\chi_m)^*$, ${\phi^\chi_3 = \phi_{nm}}$, with $u^\chi_n \equiv [\ket{u_{n\kv}}]_\chi$ the Bloch orbital coefficients, and $\phi_{nm}$ the Wilson line pseudoscalar, obtaining the tensor connection defined in the main text. Notably, for $B^{nn}_{ij}$ ($n=m$), under a gauge transformation ${u^\chi_n \rightarrow e^{\text{i}\alpha_n} u^\chi_n}$, we have: ${\phi^\chi_1 \rightarrow e^{\text{i}\alpha_n} \phi^\chi_1}$, $\phi^\chi_2 \rightarrow e^{-\text{i}\alpha_n} \phi^\chi_2$, $\phi^\chi_3 \rightarrow \phi^\chi_3 + \alpha_n$, consistently with the gerbe connection definitions of Ref.~\cite{Palumbo2019}, and consistently with the transformation, $\phi_{nn}(\kv) \rightarrow \phi_{nn}(\kv) + \alpha_n(\kv)$, as demonstrated in the main text.

Formally, a gerbe connection $B$ is defined by two-forms on charts (${U_\alpha \in \mathcal{M}}$) of a manifold $\mathcal{M}$, with curvature $\mathcal{H}$ being a~globally defined
three-form. In this work, $\mathcal{M} = \text{BZ} \cong T^3$ is the Brillouin zone manifold isomorphic to a three-torus of momenta $\textbf{k}$. The construction of a bundle gerbe with tensor connection $B$ further requires one-forms ($\mathcal{A}$) on double overlaps (${U_{\alpha \beta} = U_\alpha \cap U_\beta}$), and gluing functions ($g_{\alpha \beta \gamma}$) on triple overlaps (${U_{\alpha \beta \gamma} = U_\alpha \cap U_\beta \cap U_\gamma}$). Altogether, in terms of exterior derivative operators, $\text{d} = \text{d} k^i \frac{\partial}{\partial k^i}$, these conditions read~\cite{Murray1996},
\beq{eq:1form}
    \mathcal{A}_{\alpha \beta} + \mathcal{A}_{\beta \gamma} + \mathcal{A}_{\gamma \alpha} = -\text{i} g^{-1}_{\alpha \beta \gamma} \text{d} g_{\alpha \beta \gamma},
\eeq
\beq{eq:2form}
    B_\alpha - B_\beta = \text{d} \mathcal{A}_{\alpha \beta},
\eeq
\beq{eq:3form}
    \mathcal{H} = \text{d} B_\alpha.
\eeq
In our construction, this full data can be obtained purely from
a~Berry connection $A =A_i \text{d} k^i$ over the line bundles of Bloch vectors. {The gauge (origin) dependent line integrals of $A_i$ determine a local pseudoscalar field $\phi$ [Eq.~(5) of the main text]}. $\phi$~is a~gauge-dependent zero-form, which encodes the transition functions $g_{\alpha \beta \gamma}$ on triple overlaps $U_{\alpha \beta \gamma}$. {Formally, $g_{\alpha \beta \gamma}$ correspond to the transition functions of a second \v{C}ech cohomology~\cite{Alvarez1985}.} This follows from Eq.~\eqref{eq:1form}, as one-forms $\mathcal{A}_{\alpha \beta}$ on double overlaps $U_{\alpha \beta}$ are determined by $\phi$ obtained from integrating $A$. Explicitly, $\mathcal{A}_{\alpha \beta} \equiv \xi$, with ${\xi = \phi \text{d} \alpha}$, where $\alpha$ generates a~$U(1)$ transformation of the Berry connection: $A \rightarrow A + \text{d}\alpha$. Hence, Eq.~\eqref{eq:2form} represents a tensorial gauge transformation, $B \rightarrow B + \text{d} \xi$, introduced in the main text [Eq.~(6)]. Thus, the combination of a gauge-invariant two-form Berry flux, ${F_{ij} = \partial_j A_i - \partial_i A_j}$, with $\phi$: $B_{ij} = \phi F_{ij}$ [Eq.~\eqref{eq:2form}], determines the tensor gauge field ${B = B_{ij}~\dd k^i \wedge \dd k^j}$ two-form. ${\mathcal{H} = \mathcal{H}_{xyz}~\dd k^x \wedge \dd k^y \wedge \dd k^z}$, as a~globally defined three-form, captures the surface holonomy of tensor connection $B$, \mbox{similarly} to $F_{ij}$ capturing the holonomy of loop integrals of $A_i$. As such, $\mathcal{H}_{xyz}$ is given by the infinitesimal surface integrals of~$B_{ij}$.

{Furthermore, our momentum-space construction for the two-form gerbe connection $B$ and three-form curvature $\mathcal{H}$ is analogous to the real-space construction of the higher-order Kalb--Ramond three-form ($C$) fields and corresponding four-form field strength in higher-dimensional non-Abelian monopoles and skyrmions in quantum field theory~\cite{Kihara2005, Palumbo2022}}.

\section{Further details on Kalb--Ramond flux in quantized topological magnetoelectric response}\label{app::B} We here include more technical detail on how the gerbe connection arises in the magnetoelectric (ME) response. The magnetoelectric tensor providing for magnetization response $M_i = \a_{ij} E_j$ to electric field $E_j$ can be decomposed in terms of two terms ${\alpha_{ij} = \alpha^{\text{top}}_{ij} + \alpha^{\text{non-top}}_{ij}}$, following Ref.~\cite{Malashevich2010}, with a topological contribution central to this work,
\beq{}
    \alpha^{\text{top}}_{ij} = \frac{e^2}{2 \hbar c} \int_\text{BZ} \frac{\dd^3 \kv}{(2 \pi)^3} \sum^{\text{occ}}_{n,n'} \text{Re}~\varepsilon^{jkl} \bra{u_{n \kv}} \ket{\partial_i u_{n' \kv} } \bra{\partial_k u_{n' \kv}} \ket{\partial_l u_{n \kv} }.
\eeq
In the above, $n,n'$ run over the occupied bands (\text{occ}). We stress that the topological part provides for a $\mathbb{Z}$-quantized contribution to a magnetoelectric response. In particular, $\alpha^{\text{top}}_{ij}$ can be larger than the Chern--Simons contribution $\theta = \pi$, equal in magnitude to a contribution arising from $\mathcal{DD}_{{nn'}}=1$. In the real material context, the topological contributions are known to dominate the other, i.e., nontopological contributions~\cite{Malashevich2010_1}, to magnetoelectric coupling. On rewritting $\alpha^{\text{top}}_{ij}$ in terms of Berry connections,
\beq{}
    \alpha^{\text{top}}_{ij} = \frac{e^2}{2 \hbar c} \int_\text{BZ} \frac{\dd^3 \kv}{(2 \pi)^3} \sum^{\text{occ}}_{n,n'} \text{Re}~\varepsilon^{jkl} A^{nn'}_i \partial_k  A^{n' n}_l.
\eeq
and on combining with the definition $B^{nn'}_{ij}$ from the main text,
\begin{align}
    \nonumber \text{Tr}~\alpha^{\text{top}}_{ij} &= \frac{e^2}{2 \hbar c} \int_\text{BZ} \frac{\dd^3 \kv}{(2 \pi)^3} \sum^{\text{occ}}_{n,n'} (\partial_x B^{nn'}_{yz} + \partial_y B^{nn'}_{zx} + \partial_z B^{nn'}_{xy}) \\&= \frac{e^2}{2 \hbar c} \int_\text{BZ} \frac{\dd^3 \kv}{(2 \pi)^3} \sum^{\text{occ}}_{n,n'} \mathcal{H}^{nn'}_{xyz},
\end{align}
which demonstrates the presence of the Kalb--Ramond flux~$\mathcal{H}^{nn'}_{xyz}$, i.e., higher tensor Berry curvature, in the magnetoelectric response. Importantly, $\text{Tr}~\alpha^{\text{top}}_{ij}$ is in general $not$ just a Chern--Simons angle $\theta_{\text{CS}}$, consistently with the observations of Ref.~\cite{Ahn2022}. On~the other hand, the nontopological contribution $\alpha^{\text{non-top}}_{ij}$ reads~\cite{Malashevich2010},
\begin{align}
    \alpha^{\text{non-top}}_{ij} =& \frac{e^2}{\hbar c} \int_\text{BZ} \frac{\dd^3 \kv}{(2 \pi)^3} \sum^{\text{occ}}_{n} \sum^\text{unocc}_m\\& \nonumber \text{Re} \Bigg(\varepsilon^{jkl}  \frac{\bra{\partial_{k} u_{n\kv}} \partial_l (H + E_{n\kv}) \ket{u_{m\kv}} \bra{u_{m\kv}}\ket{ \partial_{i} u_{n\kv}}}{E_{m\kv}-E_{n\kv}} \Bigg),
\end{align}
with $m$ running over the unoccupied bands (unocc), for a~Hamiltonian $H$ and band energies $E_{n\kv}$. We also evaluate $\alpha^{\text{non-top}}_{ij}$ in the main text for comparison. We~find the contribution $\alpha^{\text{non-top}}_{ij}$ to be dominated by the topological contribution $\alpha^{\text{top}}_{ij}$ (see Fig.~3 of the main text), consistently with the findings of Ref.~\cite{Malashevich2010_1} concerning general magnetoelectric media constituted by the known materials.

\section{Further details on torsion and quantized shift response}\label{app::C} We here provide further technical details showing how the gerbe structure arises from the torsion in the quantized integrated shift effect~\cite{Jankowski2024PRL} in response to the circularly polarized light. The circular shift photoconductivity $\sigma^{\text{shift},\text{C}}_{ijk}(\om) = \text{Im}~\sigma^{\text{shift}}_{ijk}(\om)$~\cite{Ahn2020} amounts to,
\begin{align}
    \nonumber \sigma^{\text{shift},\text{C}}_{ijk}(\om) = \frac{\pi e^3}{ \hbar^2} \int_\text{BZ} \frac{\dd^3 \kv}{(2 \pi)^3}~\text{Re} \sum_{n,m} &f_{nm} \Big( C^{nm}_{kij} - (C^{nm}_{jik})^* \Big)\\& \times \delta (\om - \om_{mn}),
\end{align}
where $\om_{mn} \equiv (E_{m\kv} - E_{n\kv})/\hbar$ is an energy difference, and $f_{nm}$ are the occupation factor differences. The Hermitian connection reads $C^{nm}_{ijk} \equiv A^{nm}_i \mathcal{D}^{nm}_j A^{mn}_k$, with a covariant derivative, ${\mathcal{D}^{nm}_i \equiv \partial_i - \text{i}(A^{nn}_i-A^{mm}_i)}$~\cite{Ahn2021}. The interband torsion is defined as ${\mathcal{T}^{nm}_{ijk} \equiv C^{nm}_{ijk}  - C^{nm}_{ikj}}$, and is antisymmetric in the pair of indices ${\mathcal{T}^{nm}_{ijk} = -\mathcal{T}^{nm}_{ikj}}$ by construction. {In the above, we assume a two-state approximation for the resonant optical transitions in the considered shift current responses arising from photoexcitations.} A~spectral integration over frequencies $\om$ and an antisymmetrization over all three spatial indices obtain the left-hand side of the main text Eq.~(11) in terms of torsion, which yields the shift current sum rule originally identified in Ref.~\cite{Jankowski2024PRL}. Correspondingly, on substituting Hermitian connection and comparing with the tensor connection $B^{nm}_{ij}$ definition from the main text; in the presence of a trivialized connection term in the covariant derivative $\mathcal{D}^{nm}_i \rightarrow \partial_i$ taken in the parallel transport gauge, we have,
\begin{equation}\label{Cov:dev}
    \mathcal{T}^{nm}_{ijk} + \mathcal{T}^{nm}_{jki} + \mathcal{T}^{nm}_{kij} = \partial_i B^{nm}_{jk} + \partial_j B^{nm}_{ki} + \partial_k B^{nm}_{ij} \equiv \mathcal{H}^{nm}_{ijk}.
\end{equation}
{Eq.~\eqref{Cov:dev} follows by observing that the covariant derivative $\mathcal{D}^{nm}_j$ reduces as: ${\mathcal{D}^{nm}_j = \partial_j - \text{i}(A^{nn}_j - A^{mm}_j) \rightarrow \partial_j}$ within the Hermitian connection, ${C^{nm}_{ijk} = A^{nm}_{i}\mathcal{D}^{nm}_j A^{nm}_k}$, in the parallel transport gauge, ${A^{nn}_i = A^{mm}_i=0}$. With $\mathcal{PT}$ symmetry, this constraint can be enforced globally over the BZ, as the eigenvectors can be taken real. Under such reality condition, the diagonal Berry connection elements vanish identically, $A^{nn}_i =0$, for any band $n$.}\\

Hence, we demonstrated that the interband torsion $\mathcal{T}^{nm}_{ijk}$ amounts to the Kalb--Ramond flux $\mathcal{H}^{nm}_{ijk}$ in the momentum space. More precisely, the higher-tensor flux $\mathcal{H}^{nm}_{ijk}$ is the \text{totally} antisymmetric part of the torsion $\mathcal{T}^{nm}_{ijk}$, i.e., $\mathcal{H}^{nm}_{ijk} \propto \mathcal{T}^{nm}_{[ijk]}$, with $[ \ldots ]$ denoting an antisymmetrization in all three indices. Setting $n = m$ reduces the multiband torsion $\mathcal{T}^{nm}_{ijk}$ to an intraband torsion $\mathcal{T}^{nn}_{ijk}$, which is not present in the nonlinear shift response, but instead enters the magnetoelectric response through a flux $\mathcal{H}^{nn}_{ijk}$ central to the previous section.

\section{Homotopy classification and models of topological insulators beyond the tenfold way}\label{app::D}

In the following, we provide further details on the homotopy classification of unconventional three-dimensional topological insulators falling beyond the notion of stable equivalence underlying K-theory~\cite{Kitaev2009}.

We begin by introducing the classifying spaces $\mathcal{G}$, which capture the minimal topology of Hamiltonians $H(\kv)$ classified as maps from $d$-dimensional Brillouin zone (BZ) isomorphic to a $d$-dimensional torus, $\text{BZ} \cong T^d$, where $\textbf{k} \in \text{BZ}$. The strong topological invariants are imposed within the mapping sequence: $\text{BZ} \cong T^d \rightarrow S^d \rightarrow \mathcal{G}$, classified with homotopy groups $\pi_d[\mathcal{G}]$. The weak invariants can be included on considering more general mappings, $\text{BZ} \cong T^d \rightarrow \mathcal{G}$ classified by homotopy classes $[T^d, \mathcal{G}]$. Centrally to this work, in the following, we focus on $d=3$. 

In the absence of additional symmetries other than $\mathsf{U}(1)$ for particle/charge conservation, the corresponding classifying spaces for the Hamiltonians are complex flag manifolds, 
\beq{}
    \mathsf{Fl_{p_1,\ldots,p_k}(\mathbb{C})} \cong \frac{\mathsf{U}(N)}{\mathsf{U}(p_1) \times \ldots \times \mathsf{U}(p_k)},
\eeq
for $i = 1,2, \ldots, k$, where $i$ labels each of the $k$ band subspaces (subbundles) of rank $p_k$ equal to the number of bands therein. Each isolated subspace admits a gauge group $G_{p_i} = \mathsf{U}(p_i)$~\cite{Bouhon2020} associated with the gauge transformations in the degenerate subspace. Importantly, the presence of a spinless $(\mathcal{PT})^2=+1$ symmetry in the Hamiltonians results in an existence of a real gauge and a modified classification by real flag manifolds, $\mathsf{Fl_{p_1,\ldots,p_k}(\mathbb{R})} \cong \frac{\mathsf{O}(N)}{\mathsf{O}(p_1) \times \ldots \times \mathsf{O}(p_k)}$, on replacing each unitary gauge group with an orthogonal one: $\mathsf{U}(p_k) \rightarrow \mathsf{O}(p_k)$. Moreover, the classification can be further refined by defining the so-called orientable phases classified by oriented flag manifolds $\mathsf{Fl^+_{p_1,\ldots,p_k}(\mathbb{C})}$, $\mathsf{Fl^+_{p_1,\ldots,p_k}(\mathbb{R})}$, where the orientability condition is imposed on replacements $\mathsf{U}(p_k) \rightarrow \mathsf{SU}(p_k)$, $\mathsf{O}(p_k) \rightarrow \mathsf{SO}(p_k)$ in the corresponding definitions~\cite{Bouhon2020}.

\subsection{Two-band Hopf insulators}

We begin with two-band Hopf insulators~\cite{Moore2008, Kennedy2016}. The classifying space of a two-band Hopf insulator is a complex Grassmannian $\mathsf{Gr_{1,2}(\mathbb{C})} \equiv \mathsf{Fl_{1,1}(\mathbb{C})}$:
\beq{}
    \mathsf{Gr_{1,2}(\mathbb{C})} \cong \frac{\mathsf{U}(2)}{\mathsf{U}(1) \times \mathsf{U}(1)} \cong \frac{\mathsf{SU}(2)}{\mathsf{U}(1)} \cong S^2, 
\eeq
with the third homotopy group classifying three-dimensional bulk phases~\cite{Moore2008},
\beq{}
    \pi_3[\mathsf{Gr_{1,2}(\mathbb{C})}] \cong \pi_3[S^2] \cong \mathbb{Z}.
\eeq
$\pi_3[S^2] \cong \mathbb{Z}$ classifies distinct Hopf fibrations: $S^3 \rightarrow S^2$. In the above, we assumed an absence of weak invariants. In the presence of weak invariants, the nontrivial subdimensional Chern numbers $(C_x, C_y, C_z) =\mathbb{Z}^3$, one realizes Hopf--Chern insulator~\cite{Kennedy2016} classified by homotopy classes of maps $T^3 \rightarrow \mathsf{Gr_{1,2}(\mathbb{C})} \cong S^2$,
\begin{align}
\begin{split}
    [T^3, S^2] ={}& \Big\{(\chi; \textbf{C})\, \Big| \,  \textbf{C}=({\textit{C}_\textit{x},\textit{C}_\textit{y},\textit{C}_\textit{z}})\in \mathbb{Z}^\text{3}; \\
    &\quad \chi\in \begin{cases} \mathbb{Z} & \textbf{C} = 0\\ \mathbb{Z}_{2\gcd(\textbf{C})} & \text{\textbf{C} $\neq$ 0}\end{cases}\Bigg\}\bigg\}
\end{split},
\end{align}
where $\gcd(\textbf{C})$ is the greatest common divisor of the integer weak Chern numbers. The index $\chi$ corresponds to the strong Hopf invariant. If the Chern numbers are trivial $(C_x, C_y, C_z) = (0,0,0) = 0$, the homotopy classes reduce to the original Hopf invariant.

To retrieve the tensor monopole in the two-band Hopf insulators in the main text, we adapt the models of Refs.~\cite{Moore2008, Kennedy2016}. The studied two-band Hamiltonian $H(\kv) =\vec{d}(\kv) \cdot \pmb{\sigma}$ decomposed in the basis of Pauli matrices $\pmb{\sigma} = (\sigma_x, \sigma_y, \sigma_z)$ is generated by a vector,
\beq{eq:vector}
\vec{d}(\kv) = 
\begin{pmatrix}
    \cos k_z \sin k_x - \sin k_z \sin k_y \\
    \sin k_z \sin k_x + \cos k_z \sin k_y\\
    M - \cos k_x - \cos k_y
\end{pmatrix},
\eeq
where $M$ is a topological mass parameter yielding nontrivial Hopf invariant $\chi = \mathcal{DD}_{11} =1$, when $|M|<2$.

\subsection{N-band Hopf insulators}

The generalization of a two-band Hopf insulator to a $N$-band Hopf insulator was introduced in Ref.~\cite{Lapierre2021}. The $N$-band insulator realizes $N \geq 2$ isolated single bands constituting Bloch subbundles $\mathcal{B}_i$, each with gauge group $G_{i} = \mathsf{U(1)}$. In the isolated band limit, the classifying space reduces to,
\beq{}
    \mathsf{Fl_{1,\ldots,1}(\mathbb{C})} \cong \frac{\mathsf{U}(N)}{\mathsf{U}(1)^N}, 
\eeq
which realizes a nontrivial third homotopy group~\cite{Lapierre2021},
\beq{}
    \pi_3[\mathsf{Fl_{1,\ldots,1}(\mathbb{C})}] \cong \pi_3 [\mathsf{SU}(N)]  \cong \pi_3 [S^2] \cong \mathbb{Z},
\eeq
classifying distinct Hopf fibrations analogously to the two-band Hopf insulator.

\subsection{Three-band real Hopf insulators}

On imposing a reality condition with $\mathcal{PT}$ symmetry~\cite{Bouhon2020}, we now focus on three-band real Hopf insulators with a topologically non-trivial two-band subspace $(p_1=2)$, and a trivial third band $p_2=1$~\cite{Jankowski2024PRB}. The classifying space of a three-band real Hopf insulator is a real Grassmannian $\mathsf{Gr_{2,3}(\mathbb{R})} \equiv \mathsf{Fl_{2,1}(\mathbb{R})}$,
\beq{}
    \mathsf{Gr_{2,3}(\mathbb{R})} \cong \frac{\mathsf{O}(3)}{\mathsf{O}(2) \times \mathsf{O}(1)} \cong \mathbb{RP}^2, 
\eeq
\beq{}
    \pi_3[\mathsf{Gr_{2,3}(\mathbb{R})}] \cong \pi_3[\mathbb{RP}^2] \cong \pi_3[S^2] \cong \mathbb{Z}, 
\eeq
which defines the strong Hopf index $\mathcal{H} \in \mathbb{Z}$. The presence of weak subdimensional Euler invariants~\cite{Ahn2019, Bouhon2020, Bouhon2020ZrTe, Unal2020, bouhon2023quantumgeometry, Lim2023, Jankowski2024PRB, Wahl2025, Jankowski2025opticalprb, Jain2025} in two-dimensional BZ subtori allows for nontrivial indices $(\chi_x, \chi_y, \chi_z) \in \mathbb{Z}^3$, which when nontrivial, realize the so-called Hopf--Euler topology~\cite{Jankowski2024PRB}. The (orientable) Hopf--Euler insulators are respectively classified by homotopy classes of maps $T^3 \rightarrow \mathsf{Gr^+_{2,3}(\mathbb{R})} \cong S^2$~\cite{Jankowski2024PRB},
\begin{align}
\begin{split}
    [T^3, S^2] ={}& \Big\{( \mathcal{H} ; \pmb{\chi})\, \Big| \,  \pmb{\chi}=(\chi_\textit{x}, \chi_\textit{y}, \chi_\textit{z})\in \mathbb{Z}^\text{3}; \\
    &\quad \mathcal{H} \in \begin{cases} \mathbb{Z} & \pmb{\chi} = 0\\ \mathbb{Z}_{\gcd(\pmb{\chi})} & \text{$\pmb{\chi} \neq 0$}\end{cases}\Bigg\}\bigg\}
\end{split},
\end{align}
where $\gcd(\pmb{\chi})$ is the greatest common divisor of the integer weak Euler invariants. The index $\mathcal{H}$ corresponds to the strong Hopf invariant. When the weak Euler numbers are trivial $(\chi_x, \chi_y, \chi_z) = (0,0,0) = 0$, the homotopy classes reduce to the original real Hopf invariant.

The interband momentum-space tensor Berry connections and magnetoelectric response studied in the main text were retrieved from the model introduced in Ref.~\cite{Jankowski2024PRB}. The corresponding three-band Hamiltonian reads: $H(\kv) = 2 \vec{d}(\kv) \otimes \vec{d}^\text{T} (\kv) - \text{diag}(1.1,1,1)$ with $\vec{d}(\kv)$ defined in Eq.~\eqref{eq:vector}, and where a small diagonal perturbation was added to split the degeneracy of the occupied two-band subspace. The higher Hopf index can be generated by sending $k_z \rightarrow p k_z$, with $p \in \mathbb{N}$. The topological phase with $\mathcal{DD}_{12}=\mathcal{H}=p$ is retrieved on setting $|M| \leq 2$.

\subsection{Four-band real Hopf insulators}

Further to the three-band real Hopf insulator, we now focus on the four-band real Hopf insulators with two isolated two-band subspace $(p_1=2)$, and $(p_2=1)$~\cite{Lim2023}. The classifying space of a three-band real Hopf insulator is a real Grassmannian $\mathsf{Gr_{2,4}(\mathbb{R})} \equiv \mathsf{Fl_{2,2}(\mathbb{R})}$,
\beq{}
    \mathsf{Gr_{2,4}(\mathbb{R})} \cong \frac{\mathsf{O}(4)}{\mathsf{O}(2) \times \mathsf{O}(2)}, 
\eeq
\beq{}
    \pi_3[\mathsf{Gr_{2,4}(\mathbb{R})}] \cong \pi_3[S^2] \oplus \pi_3[S^2] \cong \mathbb{Z} \times \mathbb{Z}, 
\eeq
which defines a pair of integer real Hopf indices $(\mathcal{H}_1, \mathcal{H}_2) \cong \mathbb{Z}^2$. 

\subsection{Three-band real flag phases}

Under the partitioning of three isolated real bands, the classifying space becomes a real flag manifold distinct from the Grassmannians outlined above, defining the so-called real flag phases~\cite{Davoyan2024, Jankowski2024PRB} classified by,
\beq{}
    \mathsf{Fl_{1,1,1}(\mathbb{R})} \cong \frac{\mathsf{O}(3)}{\mathsf{O}(1) \times \mathsf{O}(1)  \times \mathsf{O}(1) }, 
\eeq
which realizes nontrivial homotopy elements,
\beq{}
    \pi_3[\mathsf{Fl_{1,1,1}(\mathbb{R})}] \cong \pi_3 [S^3] \cong \mathbb{Z}.
\eeq
The single flag index $w \in \mathbb{Z}$ can be viewed as a single three-dimensional Pontryagin index~\cite{Davoyan2024, Jankowski2024PRB}.

\subsection{Four-band real flag phases}

Analogously to the three-band case, the presence of four real bands with multiple gaps~\cite{Davoyan2024, Jankowski2024PRB}, realizes more general flag limits, beyond the previously mentioned Grassmannians. In the isolated limit of four real bands, the classifying space reads,
\beq{}
    \mathsf{Fl_{1,1,1,1}(\mathbb{R})} \cong \frac{\mathsf{O}(4)}{\mathsf{O}(1) \times \mathsf{O}(1)  \times \mathsf{O}(1)  \times \mathsf{O}(1)}, 
\eeq
which results in nontrivial homotopy group elements~\cite{Davoyan2024},
\beq{}
    \pi_3[\mathsf{Fl_{1,1,1,1}(\mathbb{R})}] \cong \pi_3 \Bigg[\frac{S^3 \times S^3}{\mathbb{Z}^2}\Bigg] \cong \pi_3 [S^3] \oplus \pi_3 [S^3] \cong \mathbb{Z} \times \mathbb{Z}.
\eeq
The nontrivial homotopy elements define a pair of flag indices $(w_L, w_R) \in \mathbb{Z}^2$ which can be viewed as isoclinic winding numbers  $\pi_3[S^3] \cong{Z}$ acting on the individual hyperspheres $S^3$~\cite{Davoyan2024, Jankowski2024PRB}.

\section{Tensor monopoles and gerbe invariants in homotopy-classified topological insulators}\label{app::E}

In the following, we detail the constructions of tensor Berry connection monopoles, i.e., topologically nontrivial configurations of momentum-space Kalb--Ramond fields $B_{ij}$ in the homotopy-classified topological insulators introduced in the previous section. Hence, we respectively map the relevant Hopf and Pontryagin indices to the $\mathcal{DD}$ invariants of nontrivial gerbe bundles constructed out of Bloch states.

\subsection{Two-band Hopf insulators}

Centrally to this work, we show that a two-band Hopf insulator realizes one momentum-space tensor monopole $B^{11}_{ij}$ with connection 2-form $B^{11}=B^{11}_{ij} \text{d}k^i \wedge \text{d}k^j$, with Kalb--Ramond flux $\mathcal{H}^{11} =\text{d}B^{11}$ determined by its exterior derivative yielding a divergence shown in the main text.
Consistently with the construction outlined in the main text, the $\mathcal{DD}_{11} \in \mathbb{Z}$ invariant reads,
\beq{}
    \mathcal{DD}_{11} = -\frac{1}{4\pi^2} \int_\text{BZ} \mathcal{H}^{11} = \chi.
\eeq

\subsection{N-band Hopf insulators}

Analogously to a two-band Hopf insulator~\cite{Moore2008}, the $N$-band Hopf insulator~\cite{Lapierre2021} with band $n$ realizing a nontrivial Abelian Chern--Simons 3-form $\mathcal{CS}^n = \varepsilon^{abc} A^n_a \partial_b A^n_c$~\cite{Lapierre2021}, hosts a single tensor monopole $B^{nn}_{ij}$ with connection 2-form $B^{nn}=B^{nn}_{ij} \text{d}k^i \wedge \text{d}k^j$ and Kalb--Ramond flux $\mathcal{H}^{nn} =\text{d}B^{nn}$. The supported nontrivial $\mathcal{DD}_{nn} \in \mathbb{Z}$ invariant reads,
\beq{}
    \mathcal{DD}_{nn} = -\frac{1}{4\pi^2} \int_\text{BZ} \mathcal{H}^{nn},
\eeq
where the explicit form of the tensor Berry connection in band $n$ is obtained by repeating the pseudoscalar $[\phi_{nn}(\kv)]$ construction outlined in the main text.

\subsection{Three-band real Hopf insulators}

We now elaborate on the three-band real Hopf insulators~\cite{Jankowski2024PRB}, which, as identified in this work, are the condensed matter realizations of an interband tensor monopole in momentum-space Kalb--Ramond fields $B^{12}_{ij}$ defining a connection 2-form $B^{12}=B^{12}_{ij} \text{d}k^i \wedge \text{d}k^j$. The higher (Berry) curvature realized in the tensor Berry connection $B^{12}_{ij}$ obtains the Kalb--Ramond flux $\mathcal{H}^{12} =\text{d}B^{12}$, consistently with the main text. The supported nontrivial interband $\mathcal{DD}_{12} \in \mathbb{Z}$ invariant reads,
\beq{}
    \mathcal{DD}_{12} = -\frac{1}{4\pi^2} \int_\text{BZ} \mathcal{H}^{12} = \mathcal{H},
\eeq
which amounts to the integer real Hopf index $\mathcal{H} \in \mathbb{Z}$ as a single momentum-space tensor monopole invariant.

\subsection{Four-band real Hopf insulators}

We now demonstrate that a four-band real Hopf insulator~\cite{Lim2023} realizes two tensor monopoles $B^{12}_{ij}$, $B^{34}_{ij}$  defining connection 2-forms $B^{12}=B^{12}_{ij} \text{d}k^i \wedge \text{d}k^j$, $B^{34}=B^{34}_{ij} \text{d}k^i \wedge \text{d}k^j$ and Kalb--Ramond fluxes $\mathcal{H}^{12} =\text{d}B^{12}$, $\mathcal{H}^{34} =\text{d}B^{34}$ under the choice of a real gauge. These determine a pair of $\mathcal{DD}$ invariants $(\mathcal{DD}_{12}, \mathcal{DD}_{34}) \in \mathbb{Z}^2$,
\beq{}
    \mathcal{DD}_{12} = -\frac{1}{4\pi^2} \int_\text{BZ} \mathcal{H}^{12},
\eeq
\beq{}
    \mathcal{DD}_{34} = -\frac{1}{4\pi^2} \int_\text{BZ} \mathcal{H}^{34},
\eeq
which combined amount to the sum of the pair of Hopf invariants, $\mathcal{DD}_{12} + \mathcal{DD}_{34}=\mathcal{H}_1+\mathcal{H}_2$.

\subsection{Three-band real flag phases}

Similarly to the three-band Hopf insulator, the three-band multigap flag phases~\cite{Jankowski2024PRB} realize a single tensor monopole in the connection $B^{12}_{ij}$. Following the construction introduced in the main text, a pair of neighboring bands defines a momentum-space Kalb--Ramond field $B^{12}_{ij}$ and a connection 2-form $B^{12}=B^{12}_{ij} \text{d}k^i \wedge \text{d}k^j$. The Kalb--Ramond flux reads $\mathcal{H}^{12} =\text{d}B^{12}$, and the interband $\mathcal{DD}$ invariant amounts to,
\beq{}
    \mathcal{DD}_{12} = -\frac{1}{4\pi^2} \int_\text{BZ} \mathcal{H}^{12} = w,
\eeq
hence, the flag index $w \in \mathbb{Z}$ is a $\mathcal{DD}$ invariant of the gerbe bundle realized in three isolated bands. 

\subsection{Four-band real flag phases}

Analogously to the four-band real Hopf insulator, the four-band flag phases~\cite{Davoyan2024} realize two tensor monopoles $B^{12}_{ij}$, $B^{34}_{ij}$, with connection 2-forms $B^{12}=B^{12}_{ij} \text{d}k^i \wedge \text{d}k^j$, $B^{34}=B^{34}_{ij} \text{d}k^i \wedge \text{d}k^j$ and Kalb--Ramond fluxes $\mathcal{H}^{12} =\text{d}B^{12}$, $\mathcal{H}^{34} =\text{d}B^{34}$ defined in the real gauge. The pair of $\mathcal{DD}$ invariants $(\mathcal{DD}_{12}, \mathcal{DD}_{34}) \in \mathbb{Z}^2$ reads,
\beq{}
    \mathcal{DD}_{12} = -\frac{1}{4\pi^2} \int_\text{BZ} \mathcal{H}^{12},
\eeq
\beq{}
    \mathcal{DD}_{34} = -\frac{1}{4\pi^2} \int_\text{BZ} \mathcal{H}^{34},
\eeq
which sums to the pair of isoclinic winding numbers~\cite{Davoyan2024} classified in the previous section, i.e., $\mathcal{DD}_{12} + \mathcal{DD}_{34}=w_L + w_R$.

\end{widetext}

\clearpage

\bibliographystyle{apsrev4-1}
\bibliography{references.bib}